\documentclass[aps,prl,showpacs,twocolumn,floats,10pt,citeautoscript,longbibliography,superscriptaddress]{revtex4-2}
\usepackage{dcolumn}
\usepackage{anyfontsize} 
\usepackage{microtype} 
\usepackage{graphicx} 
\usepackage{xspace} 
\usepackage{blindtext}
\usepackage{gensymb,bm}
\usepackage[usenames,dvipsnames]{xcolor} 

\usepackage{xparse} 

\usepackage{algorithm}
\usepackage{algpseudocode}

\usepackage{xr}
\usepackage[normalem]{ulem} 
\definecolor{darkgreen}{rgb}{0,0.5,0}
\definecolor{purple}{rgb}{0.6,0,0.5}
\definecolor{orange}{rgb}{1,0.5,0}
\definecolor{darkred}{rgb}{.7,0,0}
\definecolor{darkblue}{rgb}{0,0,.6}
\definecolor{grey}{rgb}{.6,.6,.6}
\definecolor{dimgreen}{rgb}{0.2,0.7,0.2}

\def\bk{\mathbf{k}}

\newcommand{\changed}[1]{{\color{black}{#1}}}

\newcommand{\sk}[1]{{\color{black}{#1}}}

\newcommand{\jvdx}[1]{}
\newcommand{\jvdomit}[1]{}

\usepackage[normalem]{ulem}

\usepackage{amsmath} 
\usepackage{amssymb}
\usepackage{mathrsfs} 
\usepackage{bbm}  
\renewcommand{\vec}[1]{{\boldsymbol{#1}}} 

\allowdisplaybreaks 
\usepackage{enumitem} 

\usepackage{scalefnt} 

\usepackage{tikz}
\usepackage{algorithm}
\usepackage{algpseudocode}

\usepackage[colorlinks, citecolor={blue!50!black}, urlcolor={blue!50!black}, linkcolor={red!50!black}]{hyperref}

\usepackage{scalerel,stackengine}  

\newcommand{\eLL}{{\mbox{\small$\mathscr{L}$}}}

\def\Psiodd{$|\psi_{\text{odd}}\rangle$}

\def\Psieven{$|\psi_{\text{even}}\rangle$}

\newcommand*{\ndots}{\kern-0.075em.\kern-0.05em.\kern-0.05em.}  
\newcommand*{\nidots}{.\kern-0.05em.\kern-0.05em.} 
\newcommand*{\ncdots}{\kern-0.15em\cdot\kern-0.2em\cdot\kern-0.2em\cdot\kern-0.15em}  

\newcommand{\pdag}{{\protect\vphantom{dagger}}}

\newcommand{\Nkr}{\mathcal{N}_{\mathrm{kr}}}

\def\task{TaSK}

\def\Psig{\Psi_\mathrm{g}}
\def\Eg{E_{\mathrm{g}}} 
\def\PsiI{\Psi_0}

\NewDocumentCommand{\doubleI}{O{}}{\mathbbm{1}_{#1}}
\NewDocumentCommand{\doubleIb}{O{}}{{\overline{\mathbbm{1}}_{#1}}}
\NewDocumentCommand{\doubleIk}{O{}}{\mathbbm{1}^\ks_{\! #1}}
\NewDocumentCommand{\doubleId}{O{}}{\mathbbm{1}^\ds_{\! #1}}
\NewDocumentCommand{\doubleIp}{O{}}{\mathbbm{1}^\ps_{\! #1}}
\NewDocumentCommand{\doubleV}{O{}}{\mathbbm{V}_{\! #1}}
\NewDocumentCommand{\doubleVk}{O{}}{\mathbbm{V}^\ks_{\! #1}}
\NewDocumentCommand{\doubleVd}{O{}}{\mathbbm{V}^\ds_{\! #1}}
\NewDocumentCommand{\doubleVp}{O{}}{\mathbbm{V}^\ps_{\! #1}}
\NewDocumentCommand{\doublev}{o}{{\mathbbm{v}_{#1}}}
\NewDocumentCommand{\doubleVb}{o}{{\overline{\mathbbm{V}}_{\! #1}}}
\NewDocumentCommand{\doubleVt}{o}{{\widetilde{\mathbbm{V}}_{\! #1}}}
\NewDocumentCommand{\doubleVh}{o}{\widehat{{\mathbbm{V}}_{\! #1}}}
\NewDocumentCommand{\doubleW}{o}{\mathbbm{W}_{\! #1}}
\NewDocumentCommand{\doubleWk}{o}{\mathbbm{W}^\ks_{\! #1}}
\NewDocumentCommand{\doubleWd}{o}{\mathbbm{W}^\ds_{\! #1}}
\NewDocumentCommand{\doubleWb}{o}{{\overline{\mathbbm{W}}_{\! #1}}}
\NewDocumentCommand{\doubleWt}{o}{{\widetilde{\mathbbm{V}}_{\! #1}}}
\NewDocumentCommand{\doubleWh}{o}{{\widehat{\mathbbm{V}}_{\! #1}}}

\def\ds{{\scriptscriptstyle {\rm D}}}
\def\ks{{\scriptscriptstyle {\rm K}}}
\def\ps{{\scriptscriptstyle {\rm P}}}

\def\ps{{\scriptscriptstyle {\rm P}}}

\NewDocumentCommand{\cor}{mod()}
{
	#1\IfValueTF{#2}{[#2]}{}\IfValueTF{#3}{(#3)}{}
}

\usepackage{mathtools}

\usepackage{xparse} 

\NewDocumentCommand{\xA}{
         O{fill=black}mm D<|{0} D|>{0} D//{0} O{}O{0}O{0}}  
{
	\draw [#1] (#2-00.135,#3) -- (#2,#3) -- (#2,-00.135+#3) -- cycle;
	\draw (#2-00.25-#4,#3) -- (#2-00.135,#3); 
	\draw (#2,#3) -- (#2+00.25+#5,#3); 
	\draw (#2,#3-00.135) -- (#2,#3-00.25-#6);
	\draw (#2+#8,#3+#9) node (X) {#7};
}

\NewDocumentCommand{\xAd}{
         O{fill=black}mm D<|{0} D|>{0} D//{0} O{}O{0}O{0}}  
{
	\draw [#1] (#2-00.135,#3) -- (#2,#3) -- (#2,#3+00.135) -- cycle;
	\draw (#2-00.25-#4,#3) -- (#2-00.135,#3); 
	\draw (#2,#3) -- (#2+00.25+#5,#3); 
	\draw (#2,#3+00.135) -- (#2,#3+00.25+#6);
	\draw (#2+#8,#3+#9) node (X) {#7};
}

\NewDocumentCommand{\xB}{
         O{fill=black}mm D<|{0} D|>{0} D//{0} O{}O{0}O{0}} 
{
	\draw [#1] (#2,#3) -- (#2+0.135,#3) -- (#2,-0.135+#3) -- cycle;
	\draw (#2-00.25-#4,#3) -- (#2,#3); 
	\draw (#2+0.135,#3) -- (#2+00.25+#5,#3); 
	\draw (#2,#3-0.135) -- (#2,#3-00.25-#6);
	\draw (#2+#8,#3+#9) node (X) {#7};
}

\NewDocumentCommand{\xBd}{
         O{fill=black}mm D<|{0} D|>{0} D//{0} O{}O{0}O{0}}  
{
	\draw [#1] (#2,#3) -- (#2+0.135,#3) -- (#2,#3+0.135) -- cycle;
	\draw (#2-00.25-#4,#3) -- (#2,#3); 
	\draw (#2+0.135,#3) -- (#2+00.25+#5,#3); 
	\draw (#2,#3+0.135) -- (#2,#3+00.25+#6);
	\draw (#2+#8,#3+#9) node (X) {#7};
}

\NewDocumentCommand{\xC}{
         O{fill=black}mm D<|{0} D|>{0} D//{0} O{}O{0}O{0}}  
{
	\draw [#1] (#2,#3) circle (0.065);
	\draw (#2-00.25-#4,#3) -- (#2-0.065,#3); 
	\draw (#2+0.065,#3) -- (#2+00.25+#5,#3); 
	\draw (#2,#3-0.065) -- (#2,#3-00.25-#6);
	\draw (#2+#8,#3+#9) node (X) {#7};
}

\NewDocumentCommand{\xCd}{
         O{fill=black}mm D<|{0} D|>{0} D//{0} O{}O{0}O{0}}  
{
	\draw [#1] (#2,#3) circle (0.065);
	\draw (#2-00.25-#4,#3) -- (#2-0.065,#3); 
	\draw (#2+0.065,#3) -- (#2+00.25+#5,#3); 
	\draw (#2,#3+0.065) -- (#2,#3+00.25+#6);
	\draw (#2+#8,#3+#9) node (X) {#7};
}

\NewDocumentCommand{\xW}{
         O{fill=black}mm D<|{0} D|>{0} D//{0} O{}O{0}O{0}}  
{
	\draw [#1] (#2-0.065,#3-0.065) rectangle (#2+0.065,#3+0.065);
	\draw (#2-00.25-#4,#3) -- (#2-0.065,#3); 
	\draw (#2+0.065,#3) -- (#2+00.25+#5,#3); 
	\draw (#2,#3-0.065) -- (#2,#3-00.25-#6);
	\draw (#2,#3+0.065) -- (#2,#3+00.25+#6);
	\draw (#2+#8,#3+#9) node (X) {#7};
}

\NewDocumentCommand{\lcurl}{mmm  O{}O{0}O{0} D<>{0}}   
{
	\draw (#1,#3) edge[out=180,in=-90] (#1-0.1+#7,#2*0.5+#3*0.5);
	\draw (#1-0.1+#7,#2*0.5+#3*0.5) edge[out=90,in=180] (#1,#2);
	\draw (#1+#5,0.5*#3+0.5*2+#6) node (X) {#4};
}

\NewDocumentCommand{\rcurl}{mmm  O{}O{0}O{0} D<>{0}}   
{
	\draw (#1,#3) edge[out=0,in=-90] (#1+0.1+#7,#2*0.5+#3*0.5);
	\draw (#1+0.1+#7,#2*0.5+#3*0.5) edge[out=90,in=0] (#1,#2);
	\draw (#1+#5,0.5*#2+0.5*#3+#6) node (X) {#4};
}

\NewDocumentCommand{\effL}{O{fill=black} mmm  O{$\,$}O{0}O{0} D<>{0}}   
{
	\draw (#2,#4) edge[out=180,in=-90] (#2-00.2+#8,#3*0.5+#4*0.5-0.1);
	\draw (#2-00.2+#8,#3*0.5+#4*0.5+0.1) edge[out=90,in=180] (#2,#3);
	\draw [#1](#2-00.2+#8,#3*0.5+#4*0.5+0.1) -- 
	(#2-00.2+#8,#3*0.5+#4*0.5-0.1) -- (#2-00.1+#8,#3*0.5+#4*0.5) -- cycle;
	\draw (#2-00.2+#8,#3*0.5+#4*0.5) -- (#2,#3*0.5+#4*0.5);
	\draw (#2+#6,#4+#7) node (X) {#5};
}

\NewDocumentCommand{\effR}{O{fill=black} mmm  O{$\,$}O{0}O{0} D<>{0}}   
{
	\draw (#2,#4) edge[out=0,in=-90] (#2+00.2+#8,#3*0.5+#4*0.5-0.1);
	\draw (#2+00.2+#8,#3*0.5+#4*0.5+0.1) edge[out=90,in=0] (#2,#3);
	\draw [#1] (#2+00.2+#8,#3*0.5+#4*0.5+0.1) -- 
	(#2+00.2+#8,#3*0.5+#4*0.5-0.1) -- (#2+00.1+#8,#3*0.5+#4*0.5) -- cycle;
	\draw (#2+00.2+#8,#3*0.5+#4*0.5) -- (#2,#3*0.5+#4*0.5);
	\draw (#2+#6,#4+#7) node (X) {#5};
}

\makeatletter
\def\maketitle{
\@author@finish
\title@column\titleblock@produce
\suppressfloats[t]}
\makeatother

\def\maintitle{ Revisiting the $J_1$-$J_2$ Heisenberg Model on a Triangular Lattice: \\
Quasidegenerate Ground States and Phase Competition}

\begin{document} 

\title{\maintitle}

\author{Oleksandra Kovalska}
\altaffiliation{OK and EPF contributed equally to this work. Contact author: o.kovalska@lmu.de}
\affiliation{Arnold Sommerfeld Center for Theoretical Physics, 
Center for NanoScience,\looseness=-1\,  and 
Munich Center for \\ Quantum Science and Technology,\looseness=-2\, 
Ludwig-Maximilians-Universit\"at M\"unchen, 80333 Munich, Germany}

\author{Ester Pagès Fontanella}
\altaffiliation{OK and EPF contributed equally to this work. Contact author: o.kovalska@lmu.de}
\affiliation{Arnold Sommerfeld Center for Theoretical Physics, 
Center for NanoScience,\looseness=-1\,  and 
Munich Center for \\ Quantum Science and Technology,\looseness=-2\, 
Ludwig-Maximilians-Universit\"at M\"unchen, 80333 Munich, Germany}

\author{Benedikt Schneider}
\affiliation{Arnold Sommerfeld Center for Theoretical Physics, 
Center for NanoScience,\looseness=-1\,  and 
Munich Center for \\ Quantum Science and Technology,\looseness=-2\, 
Ludwig-Maximilians-Universit\"at M\"unchen, 80333 Munich, Germany}

\author{Hong-Hao Tu}
\affiliation{Arnold Sommerfeld Center for Theoretical Physics, 
Center for NanoScience,\looseness=-1\,  and 
Munich Center for \\ Quantum Science and Technology,\looseness=-2\, 
Ludwig-Maximilians-Universit\"at M\"unchen, 80333 Munich, Germany}

\author{Jan von Delft}
\affiliation{Arnold Sommerfeld Center for Theoretical Physics, 
Center for NanoScience,\looseness=-1\,  and 
Munich Center for \\ Quantum Science and Technology,\looseness=-2\, 
Ludwig-Maximilians-Universit\"at M\"unchen, 80333 Munich, Germany}

\date{\today}

\begin{abstract}
It is generally believed that the spin-$\tfrac{1}{2}$ triangular-lattice $J_1$-$J_2$ Heisenberg model hosts a quantum spin liquid in the intermediate regime between the $120\degree$ and stripe ordered phases. Density matrix renormalization group studies on cylinders have consistently found two nearly degenerate ground states, commonly interpreted as distinct topological sectors. Using state-of-the-art matrix product state simulations on YC6 cylinders, we compare the static and dynamical properties of these two sectors at $J_2/J_1 = 0.125$. Noticeable differences appear already in static correlations; moreover, high-resolution dynamical structure factors reveal qualitatively distinct low-energy excitations. These results suggest that the two ground states cannot be understood as merely topologically distinct sectors of a gapped $\mathbb{Z}_2$ spin liquid.
\end{abstract}

\maketitle

\textit{Introduction.---}
The spin-$\tfrac{1}{2}$ antiferromagnetic Heisenberg model on a triangular lattice~(TLHAF) has long served as a paradigmatic platform for studying frustrated quantum magnetism. In his pioneering work, Anderson argued that the TLHAF with only nearest-neighbor~(NN) coupling could have a magnetically disordered ground state due to geometric frustration~\cite{Anderson1973}. Subsequent numerical studies, however, established that the ground state of the TLHAF with NN coupling instead has a coplanar $120\degree$ magnetic long-range order~\cite{Bernu1994,Capriotti1999,White2007,Mezzacapo2010,Li2022,Huang2024}.

Introducing further competing interactions can enhance quantum fluctuations and potentially destabilize the $120\degree$ magnetic order. In particular, adding a next-nearest-neighbor~(NNN) antiferromagnetic coupling leads to the so-called $J_1$-$J_2$ Heisenberg model:
\begin{align}
    H = J_1 \sum_{\langle i,j\rangle}\vec{\mathrm{S}_i}\cdot\vec{\mathrm{S}_j} + J_2 \sum_{\langle\!\langle i,j\rangle\!\rangle}\vec{\mathrm{S}_i}\cdot\vec{\mathrm{S}_j},
    \label{eq:HeisenbergHamiltonian}
\end{align}
where $\langle i,j\rangle$ and $\langle\!\langle i,j\rangle\!\rangle$ denote NN and NNN pairs on the triangular lattice, respectively. For small $J_2/J_1$, the $120\degree$ magnetic order remains robust, whereas at sufficiently large $J_2/J_1$, the system develops a collinear stripe order~\cite{Jolicoeur1990,Chubukov1992}. Between these two magnetically ordered phases, extensive numerical studies~\cite{Manuel1999,Mishmash2013,Kaneko2014,Li2015,Zhu2015,Hu2015,Bishop2015,Iqbal2016,Hu2019,Ferrari2019,Gong2019,Tang2022,Jiang2023,Wietek2024} have identified an intermediate nonmagnetic regime in the approximate range $0.07\lesssim J_2/J_1\lesssim 0.15$, exhibiting signatures of a quantum spin liquid (QSL)~\cite{Wen2002,Balents2010,Zhou2017,Kivelson2023}. A schematic phase diagram is shown in Fig.~\ref{fig:QSL_phase_diagram}.

\begin{figure}[t]
 \includegraphics[width=\linewidth]{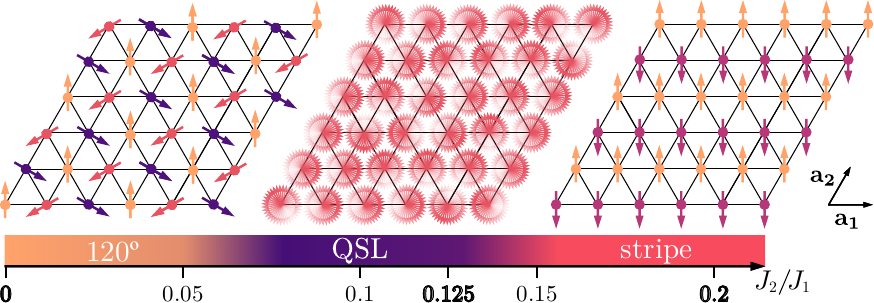} \vspace{-8mm}
\caption{
A schematic phase diagram of the $J_1$-$J_2$ Heisenberg model on a triangular lattice as a function of $J_2/J_1$, as conjectured from previous numerical studies~\cite{Manuel1999,Mishmash2013,Kaneko2014,Li2015,Zhu2015,Hu2015,Bishop2015,Iqbal2016,Hu2019,Ferrari2019,Gong2019,Tang2022,Jiang2023,Wietek2024}. For $J_2 = 0$ the model exhibits coplanar $120\degree$ order, transitioning into the QSL candidate region around $J_2/J_1 \approx 0.07$, followed by a stripe-ordered phase at larger values of coupling. }
\label{fig:QSL_phase_diagram} 
\vspace{-6mm}
 \end{figure}

The nature of this putative QSL remains under active debate. Two prominent candidates have been proposed: a gapped $\mathbb{Z}_2$ spin liquid~\cite{Zhu2015,Lu2016,Saadatmand2016,Bauer2017,Jiang2023} and a gapless U(1) Dirac spin liquid~\cite{Iqbal2016,Song2019,Hu2019,Song2020,Drescher2023,Sherman2023,Wietek2024,Seifert2024,Budaraju2025,Willsher2025,Drescher2025}. Since QSLs do not break global  symmetries and lack local order parameters, distinguishing these scenarios using numerical methods is inherently challenging. Interestingly, several independent density matrix renormalization group (DMRG)~\cite{White1992,Schollwoeck2011} studies on finite-width cylinders have consistently reported the presence of two nearly degenerate ground states~\cite{Zhu2015,Hu2019,Drescher2023}. This observation has often been interpreted as being consistent with the presence of topological (or flux) sector structure~\cite{Hu2019,Drescher2023}: on a finite cylinder, different topological sectors may manifest as nearly degenerate ground states separated by small energy splittings. These states are commonly referred to as even and odd sectors~\cite{Hu2019,Drescher2023}. In a gapped topological phase, such sectors are expected to become exactly degenerate in the thermodynamic limit and indistinguishable by any local observable in the bulk. However, similar near degeneracies may also arise in gapless spin liquids or from finite-size effects. Therefore, whether the observed near degeneracy unambiguously signals topological order remains unsettled.

In this work, we employ large-scale matrix product state (MPS) simulations to shed light on the nature of the even- and odd-sector ground states on a width-6
cylinder at $J_2/J_1 = 0.125$. The latter is a representative point within the proposed nonmagnetic regime and coincides with the transition point of the corresponding classical model~\cite{Jolicoeur1990}. We demonstrate that the two sectors exhibit clear differences in local observables. Qualitative differences appear already in static correlations: (i) the equal-time structure factor exhibits substantially different intensity profiles across the Brillouin zone, and (ii) the NN bond correlations pattern in the odd sector is noticeably more isotropic.
Most importantly, their low-energy dynamical responses are drastically different. While the even sector exhibits comparable spectral intensity at the $K$ and $M$ points in the Brillouin zone, the odd sector concentrates most of its spectral weight at the $K$ point, reminiscent of the proximate $120\degree$ ordered phase. These results indicate that the two ground states cannot be understood as arising from the topological sectors of a gapped $\mathbb{Z}_2$ spin liquid. Instead, our findings call for a careful reassessment of the origin of the two-sector structure in this parameter regime.

\begin{figure}[b]
\includegraphics[width=\linewidth]{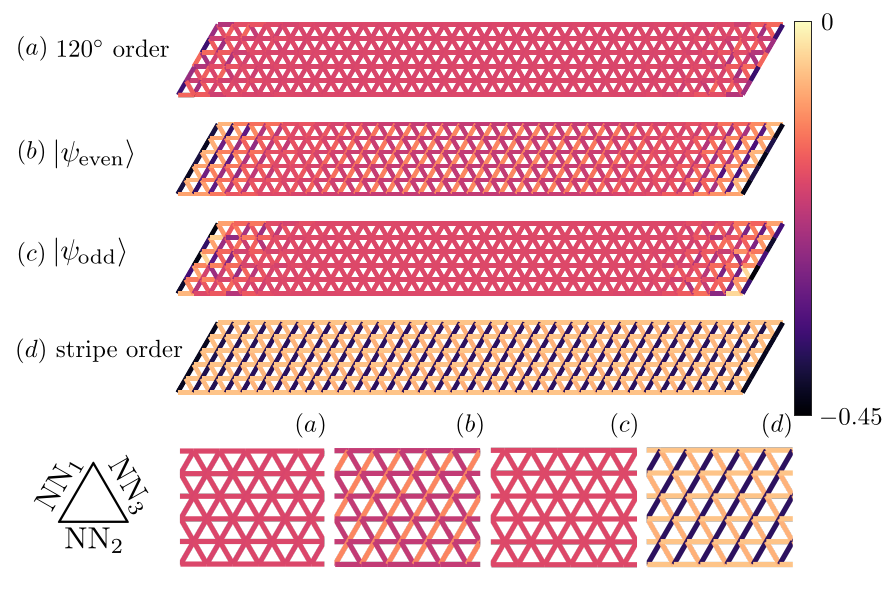} 
\caption{Nearest-neighbor spin-spin correlation $S_{\langle i,j\rangle}$ on the YC6-36 cylinder computed for (a) $120\degree$ ordered state at $J_2 = 0$, (b) even- and (c) odd-sector states at $J_2/J_1 = 0.125$, and (d) stripe-ordered state at $J_2/J_1 = 0.2$. Bond strengths are represented via a colormap, with the same scale applied globally across all panels to allow direct comparison.\sk{ The bottom row contains insets of the four central rings of each cylinder respectively.} }
\label{fig:QSL_bonds} 
\vspace{-5mm}
\end{figure} 

\textit{Grounds states on a cylinder.---} 
We study the Hamiltonian~\eqref{eq:HeisenbergHamiltonian} on a triangular lattice wrapped onto a finite cylinder. We denote by $L_y$ ($L_x$) the number of spins along the circumference (length) of the cylinder. Throughout this work we adopt the YC geometry~\cite{Zhu2015}, imposing a periodic boundary condition along $\mathbf{a_2}$ by identifying sites at $\vec{\mathrm{r}}$ and $\vec{\mathrm{r}} + L_y \vec{\mathrm{a}_2}$, with primitive vectors $\mathbf{a_1} = (1,0)$ and $\mathbf{a_2}=(1/2, \sqrt{3}/2)$ (see Fig.~\ref{fig:QSL_phase_diagram}). Unless stated otherwise, the calculations are performed on YC6-36 cylinders, corresponding to $L_y = 6$ and $L_x = 36$. In the following, we set the energy unit as $J_1 = 1$.

The ground-state MPSs are obtained using single-site DMRG with controlled bond expansion~\cite{Gleis2023}. We use the QSpace tensor library~\cite{Weichselbaum2012,Weichselbaum2020,Weichselbaum2024} to exploit the non-Abelian $\mathrm{SU}(2)$ symmetry of the Hamiltonian~\eqref{eq:HeisenbergHamiltonian}. The simulations are performed until the targeted MPS reaches at least $D^*=3000$
kept SU(2) multiplets, after which it is compressed down to $D^*=1000$ for subsequent use. We find this strategy to be optimal for the TLHAF, particularly in the QSL regime where convergence is hindered by substantial frustration and entanglement.    

Previous studies of the TLHAF on cylindrical geometry have consistently found two nearly degenerate ground states in the QSL candidate phase~\cite{Hu2015,Zhu2015,Saadatmand2016,Gong2017,Drescher2023}, whose properties, at large, remained unclear. When considering infinitely long cylinders, for example, by means of iDMRG, direct simulations yielded only the higher-energy state (the even sector)~\cite{Drescher2023}. The ground state in the odd sector was then reached through flux insertion~\cite{Hu2019,Drescher2023}, which explicitly breaks the SU(2) symmetry.

It is not obvious how to systematically obtain both states while maintaining the $\mathrm{SU}(2)$ symmetry of the microscopic Hamiltonian~\eqref{eq:HeisenbergHamiltonian}. In their DMRG study of finite-size even-$L_y$ cylinders, Zhu and White~\cite{Zhu2015} reported that, for sufficiently large $L_x$ and at a certain number of kept states, the MPS ground state dynamically falls into the odd sector. Similarly, when running DMRG in the QSL regime for YC6-36, we observe a sudden drop in energy while performing the sweeping routine for $D^* = 2048$ (see Sec.~\ref{sec:SM_DMRG} of the Supplemental Material~\cite{supplement}). This behavior is reproducible across multiple independent DMRG simulations with randomized initial states, demonstrating its robustness. Conversely, for shorter cylinders ($L_x = 18$), our ground-state optimization routine does not exhibit any such abrupt changes. 

Throughout this work, we will refer to the states before and after the drop as \Psieven \ and \Psiodd, respectively. While it is not immediately clear how they are related, we note that the energy of \Psiodd \ is lower by $1.1\%$, and the nature of NN spin-spin correlations changes drastically across the drop (see Fig.~\ref{fig:QSL_bonds}). Moreover, we find that the two states are nearly orthogonal, with a fidelity $\mathcal{F} = |\langle\psi_{\text{even}}|\psi_{\text{odd}}\rangle|$ of order $\sim 10^{-4}$. Motivated by these observations, in the following sections, we examine the static and dynamical properties of the even and odd sectors in greater detail, comparing them with each other and with results obtained for the ordered phases.

 \begin{figure*}[t]
\includegraphics[width=\textwidth]{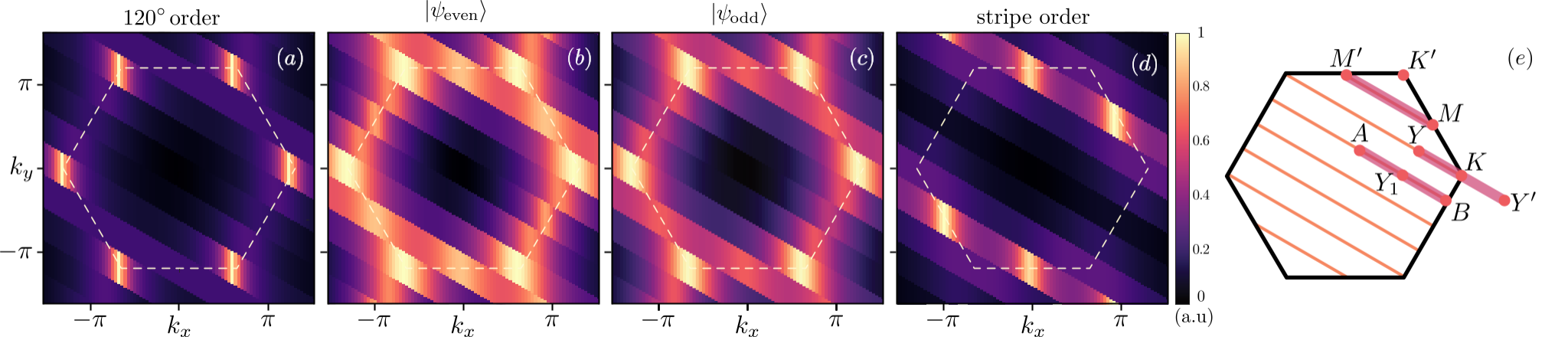} 
\vspace{-7mm}
 \caption{The equal-time structure factor $\chi(\vec{\mathrm{k}})$ extracted from ground-state MPSs of the YC6-36 cylinder. Panels (a) and (d) correspond to $J_2 = 0$ ($120\degree$ order) and $J_2/J_1 = 0.2$ (stripe order), respectively. Results for even (b) and odd (c) ground states in the QSL candidate phase are plotted side-by-side for comparison. 
Each panel is normalized independently to its own maximum intensity to highlight the relative distribution of spectral weight across different coupling regimes.
The dashed white lines indicate the boundaries of the first Brillouin zone, which we also plot separately in panel (e), marking high symmetry points and the cuts considered in our DSF computations (see Fig.~\ref{fig:QSL_DSF}). 
\label{fig:QSL_static} }
\vspace{-3mm}
\end{figure*}

\textit{Static correlations.---} 
Figure~\ref{fig:QSL_static} displays the equal-time structure factor~(ETSF)
\begin{align}
    \chi(\vec{\mathrm{k}}) = \frac{1}{\eLL} \sum_{i,j} e^{-i\vec{\mathrm{k}}\cdot(\vec{\mathrm{r}_i}-\vec{\mathrm{r}_j})} \langle \vec{\mathrm{S}_i}\cdot\vec{\mathrm{S}_j} \rangle 
    \vspace{-2mm}
\end{align} obtained from ground-state MPSs on the YC6-36 cylinder. Here, $\eLL = L_y \times L_x$ denotes the total number of sites in the system.  

In the $120\degree$ ordered phase at $J_2 = 0$ [Fig.~\ref{fig:QSL_static}(a)], the ETSF shows sharp isolated maxima at the corners of the Brillouin zone, corresponding to $K$ and $K'$ points. This is consistent with the presence of long-range magnetic order in the ground state, which manifests through high-intensity Bragg peaks. In the stripe-ordered phase at $J_2/J_1 = 0.2$ [Fig.~\ref{fig:QSL_static}(d)], sharp features instead emerge at the $M$ and $M'$ points, with little to no intensity across the remainder of the Brillouin zone, characteristic of stripe formation.

For the study of the QSL candidate phase, we set $J_2/J_1 = 0.125$, which lies within the putative QSL region (see also Fig.~\ref{fig:QSL_phase_diagram}). In contrast to magnetically ordered states, U(1) Dirac QSLs are expected to host a broad distribution of correlations across the Brillouin zone, up to anisotropies induced by the cylindrical geometry. This scenario is most consistent with the even-sector ETSF [Fig.~\ref{fig:QSL_static}(b)], where a significant softening of the sharp features is observed at the $K$ and $M$ points, compared to the ordered phases. Moreover, these points exhibit comparable intensity, which is compatible with expectations for a U(1) Dirac QSL.

For the odd-sector ETSF [Fig.~\ref{fig:QSL_static}(c)], the correlations likewise become broader near the corners of the Brillouin zone, albeit less prominently than in the even sector. The maxima at the $M$ points are less pronounced in comparison, in particular those located at $\pm (\pi,\pi/\sqrt{3})$.

We now turn to the NN spin-spin correlations $ S_{\langle i,j\rangle} = \langle \vec{\mathrm{S}_i}\cdot\vec{\mathrm{S}_j} \rangle $, shown in Fig.~\ref{fig:QSL_bonds}. In the bulk of the cylinder, the bond strength patterns for \Psieven\ [Fig.~\ref{fig:QSL_bonds}(b)] and \Psiodd\ [Fig.~\ref{fig:QSL_bonds}(c)] are strikingly different. The former exhibits a staggering pattern, similar to the one observed in the stripe-ordered phase [Fig.~\ref{fig:QSL_bonds}(d)]. However, the difference in amplitude between the three distinct bonds is much smaller, with the bond strength for $\mathrm{NN_1}$ differing significantly from $\mathrm{NN_2}$ and $\mathrm{NN_3}$ (see Tab.~\ref{tab:QSL_table}). Conversely, for the odd sector, the correlations appear to be isotropic and close in value to those obtained in the $120\degree$ ordered phase [Fig.~\ref{fig:QSL_bonds}(a)], with substantial differences arising only at edges.

\renewcommand{\arraystretch}{1.125}
\setlength{\tabcolsep}{4pt}
\begin{table}[h]
\centering
\begin{tabular}{|c|c|c|c|c|}
\hline
 & $120^\circ$ order 
 & \Psieven
 & \Psiodd
 & stripe order \\
\hline
NN$_1$ & $-0.1874$  & $-0.1124$ & $-0.1804$ & $-0.3763$ \\
\hline
NN$_2$ & $-0.1839$ & $-0.2113$ & $-0.1821$ & $-0.0542$ \\
\hline
NN$_3$ & $-0.1837$ & $-0.2111$ & $-0.1821$ & $-0.0733$ \\
\hline
\end{tabular}
\caption{Nearest-neighbor spin-spin correlations evaluated on the central ring of the YC6-36 cylinder for four different ground states at $J_2/J_1 = 0$, $0.125$, and $0.2$, respectively. $\mathrm{NN_i}$ ($i=1,2,3$) denotes a distinct pair of neighboring spins on a triangular lattice, as depicted in Fig.~\ref{fig:QSL_bonds}.}
\label{tab:QSL_table}
\vspace{-4mm}
\end{table}

The above analysis of the static properties of the four states suggests that the two QSL ground states exhibit notable differences in their local observables. In particular, \Psiodd \ appears to share certain features with the ordered phase at $J_2 = 0$. 

\textit{Dynamic correlations.---}
To further elucidate the differences between the states discussed above, we compute the dynamical structure factor~(DSF)
\begin{align}
    S(\vec{\mathrm{k}},\omega)
    = \langle \Psig |\vec{\mathrm{S}}^{\dag}_\bk\, \delta(\omega - H + 
    \Eg)\, \vec{\mathrm{S}}_\bk^{\pdag} |\Psig \rangle,
    \label{eq:SpectralFunction1}
\end{align}
using the tangent space Krylov~(\task) method~\cite{Kovalska2025} with DMRG ground states as input.

In this approach, we approximate resolvents by projecting the Hamiltonian to the tangent space of the ground-state MPS, denoted by $\doubleV^{1\perp}$. The explicit vector-space structure of $\doubleV^{1\perp}$ allows us to construct a Krylov basis through iterative application of the projected Hamiltonian $ H^{1\perp}$ to the initial state $ |\PsiI\rangle = \vec{\mathrm{S}}_\bk^{\pdag} |\Psig \rangle$
 \begin{align}
\mathcal{K}\left(\PsiI\right) 
&= \mathrm{span}\bigl\{|\PsiI\rangle,H^{1\perp}|\PsiI\rangle, \dots, \bigl(H^{1\perp})^{\Nkr}|\PsiI\rangle\bigr\} \, ,
\label{eq:Krylov_space}
\end{align}
where $\Nkr$ is the number of Krylov steps. This is achieved via a stable realization of the Lanczos algorithm, meaning that at every iteration, all vectors are mutually orthogonal down to numerical precision. In comparison to time-evolution-based tensor network methods, when computing dynamical correlators, \task \ achieves high resolution at low frequencies much more efficiently.

The real discrete frequencies and spectral weights $\{\omega_\alpha,S_{\alpha}\}$, approximating $S(\vec{\mathrm{k}},\omega)$, are extracted from the matrix representation of $ H^{1\perp}$ in the Krylov basis. To obtain smooth spectra, we broaden them using continued fraction expansion scheme, which conserves the total spectral weight and the lowest frequency moments of the DSF (see Sec.~S-2 B of the Supplemental Material of Ref.~\onlinecite{Kovalska2025}). The discrete data and broadened curves corresponding to other choices of the broadening parameters can be found in the Supplemental Material~\cite{supplement}.

Figure~\ref{fig:QSL_DSF} shows the dynamical structure factor computed with \task \ for the $120\degree$ ordered phase [Fig.~\ref{fig:QSL_DSF}(a)] and QSL candidate phase [Fig.~\ref{fig:QSL_DSF}(b,c)]. We focus on two momentum cuts containing $K$ and $M$ points [see Fig.~\ref{fig:QSL_static}(e)]. To enable a fair comparison of spectral intensities, each row is normalized to its maximum value $S_{\mathrm{max}}(\vec{\mathrm{k}},\omega)$ among all considered momentum points. For line cuts through the $K$ and $M$ point as well as a discussion of results obtained for the $A-B$ path, including the conjectured magnon minimum at the $Y_1$ point~\cite{Verresen2019}, see
the Supplemental Material~\cite{supplement}. 

In the $120\degree$ ordered phase [Fig.~\ref{fig:QSL_DSF}(a)], the DSF exhibits a sharp magnon Goldstone mode, gapless at the $K$ point up to finite-size effects imposed by cylindrical geometry. At the centers of the Brillouin zone edges ($M$ and $M'$ points), the spectral weight is significantly suppressed and has roton-like minima (not visible in this normalization, see also Fig.~\ref{fig:DSF_old} in the Supplemental Material~\cite{supplement}). Overall, our results at $J_2=0$ are in conclusive agreement with previous theoretical and numerical studies~\cite{Ghioldi2018,Ferrari2019,Chi2022,Drescher2023}.

Comparing the DSFs of \Psieven \ [Fig.~\ref{fig:QSL_DSF}(b)] and \Psiodd \ [Fig.~\ref{fig:QSL_DSF}(c)] in the QSL regime, we find that the two states exhibit similar spectral profiles, distinct from that of the $120\degree$ ordered state. However, significant differences arise in the distribution of spectral weight: for \sk{\Psiodd}, it is mostly concentrated at the $K$ point, whereas for \sk{\Psieven} it is distributed more uniformly throughout the Brillouin zone, peaking at the $K$ and $M$ points with comparable intensity. In agreement with the ETSF results, these observations are consistent with the even sector potentially being a U(1) Dirac QSL~\cite{Ferrari2019,Drescher2023}, while the nature of the odd sector remains, at large, unresolved, albeit exhibiting signatures of a proximate $120\degree$ state. \sk{We also note that for the Dirac QSL the DSF should be gapless at the M points (since the spinon Dirac cones are expected to occur at $\pm Q = \pm (\pi/2,\pi/2\sqrt{3}))$ \cite{Drescher2023,Drescher2025}. The finite gap, observed in Figs.~\ref{fig:QSL_DSF}(b) and \ref{fig:QSL_DSF}(c) is attributed to the finite circumference of the cylinder, and we expect it to close in the  $L_y \rightarrow \infty$ limit.}

\begin{figure}[t]
\includegraphics[width=\linewidth]{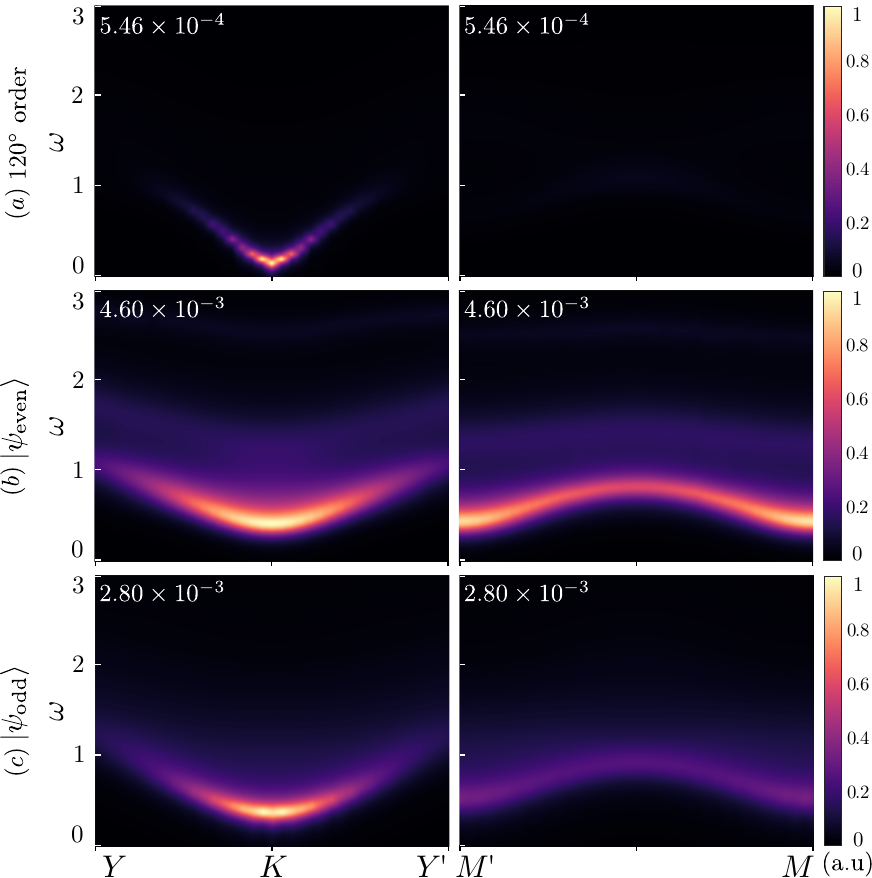} 
\caption{The dynamical structure factor computed for (a) the $120\degree$ ordered phase at $J_2 = 0$  and (b,c) the putative QSL phase at $J_2/J_1 = 0.125$ along the momentum cuts indicated in Fig.~\ref{fig:QSL_static}(e). The normalization factor
$1/S_{\mathrm{max}}(\vec{\mathrm{k}},\omega)$ is provided in the top left corner of each panel. }
\label{fig:QSL_DSF} 
\vspace{-5mm}
 \end{figure} 

\textit{Summary and outlook.---} Our work provides compelling evidence that the ``even'' and ``odd'' sectors of the TLHAF model, thus far thought to be only topologically distinct, exhibit substantial differences in their locally observable properties on the YC6 cylinder. This result rules out the mere emergence of quasi-degenerate ground states as a signature of a gapped $\mathbb{Z}_2$ QSL. Instead, we find that the even-sector ground state could be interpreted as the U(1) Dirac QSL, while the odd-sector ground state, which appears to be the stable configuration for this geometry, shares many low-energy features with the magnetically ordered phase at $J_2=0$. The latter finding has also been reported in recent time-evolution studies of the same model~\cite{Drescher2023, Drescher2025}, which focused on the odd sector. 

Several questions, therefore, remain open: (i) Is the true ground state always in the odd sector -- i.e., the state with more isotropic correlations -- or could the two states trade places as system size and geometry are varied? (ii) Does the odd-sector ground state represent a QSL, whose nature is different from U(1) Dirac, or is it a proximate $120\degree$ state? Providing definitive answers will require systematic, large-scale DMRG studies on wider cylinders and different geometries. 

More generally, unambiguous identification of spectral features in the QSL regime remains
a major challenge, particularly in tensor-network studies. On this front, additional insights could be gained by extracting dynamical correlations from MPS representations of parton wave functions~\cite{Wu2020,Jin2020,Aghaei2020,Petrica2021,Jin2025,Hille2025} corresponding to candidate QSL Ansätze. Such calculations can be readily carried out within the \task\ framework. Exploring this direction in detail, however, lies beyond the scope of the present work and is left for future study.

\textit{Note added.---}
During the finalization of this manuscript, we became aware of \changed{Ref.~}%
\cite{Jiang2026}, 
\changed{which} 
likewise compares the static and dynamical properties of two nearly degenerate ground states of the TLHAF in the intermediate $J_2/J_1$ regime, \changed{on cylinders of width 6 to 9}. While both their and our works identify two nearly degenerate states and discuss their distinct features, the present work emphasizes a complementary perspective based on SU(2)-symmetric MPS simulations paired with the tangent space Krylov approach, which enables a direct evaluation of the dynamical structure factor within the MPS tangent-space framework. 

\changed{\textit{Data availability.---} The simulation data including plotting scripts are available via the zenodo repository at Ref.~\cite{data}.}

\begin{acknowledgments}

\textit{Acknowledgments.---} 
We thank Eric Andrade, Carlo de Castro Bellinati, Andreas Gleis, Jheng-Wei Li, Wei Li, Bruce Normand, Frank Pollmann, Marc Ritter, and Wei Zhu for helpful discussions, and Markus Drescher for sharing the quasiparticle ansatz data.
This work was supported in part by the Deutsche Forschungsgemeinschaft under grants INST 86/1885-1 FUGG,  LE 3883/2-2 and 
Germany's Excellence Strategy EXC-2111 (Project No.~390814868). It is part of the  Munich Quantum Valley, supported by the Bavarian state government with funds from the Hightech Agenda Bayern Plus. 
The National Science Foundation supported JvD in part under PHY-1748958. 
\end{acknowledgments}

\raggedbottom 

\vspace{-1cm}

\bibliography{TLHAF_paper}

\clearpage

\title{Supplemental material: \\ \maintitle}

\date{\today}
\maketitle

\setcounter{secnumdepth}{2} 
\renewcommand{\thefigure}{S-\arabic{figure}}
\setcounter{figure}{0}
\setcounter{section}{0}
\setcounter{equation}{0}
\renewcommand{\thesection}{S-\arabic{section}}
\renewcommand{\theequation}{S\arabic{equation}}

\section{DMRG Sweeping Routine}
\label{sec:SM_DMRG}

\begin{figure}[h]
 \includegraphics[width=\linewidth]{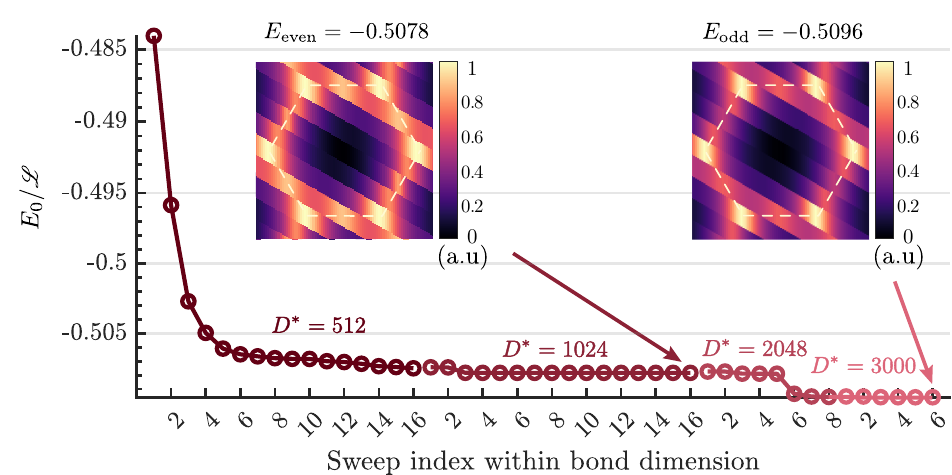} \vspace{-7mm}
\caption{DMRG sweeping protocol for the TLHAF at $J_2/J_1 = 0.125$. The two insets show equal-time structure factors for the MPS before and after the drop observed around the fifth sweep at $D^* = 2048$.
}
\label{fig:QSL_DMRG} 
\vspace{-3mm}
 \end{figure}

Figure~\ref{fig:QSL_DMRG} follows our DMRG sweeping routine for the YC6-36 cylinder. In addition to the controlled bond expansion scheme~\cite{Gleis2023}, we introduce a DMRG3S-like mixing parameter~\cite{Hubig2015,Gleis2025b} during the first two half-sweeps whenever the number of SU(2) multiplets $D^*$ is increased. For $D^* > 512$, we perform at least eight full sweeps. Within this protocol, we observe a steady convergence trend up until $D^* = 2048$, where a pronounced drop in the variational energy density occurs at the fifth sweep. 

The energy per site in the even sector $E_{\text{even}}$ right before the drop is higher than in the odd sector $E_{\text{odd}}$ by $1.1\%$. Moreover, while the fidelity $ \mathcal{F} = |\langle\psi_{512}|\psi_{1024}\rangle|$ between the states at $D^* = 512$ and $D^* = 1024$  is larger than $89\%$, its value drastically drops to  $10^{-4}$ for $|\langle\psi_{1024}|\psi_{3000}\rangle|$. This behavior is reproducible across independent DMRG runs, where we initiate the state with randomized tensor entries within the fixed symmetry sector. For substantially shorter $L_x$, e.g., when considering YC6-18, the optimization routine proceeds without any such ``anomalies", in agreement with general observations from Ref.~\onlinecite{Zhu2015}.

For the odd-sector ground state, all correlations presented in the main text are extracted from $|\phi_{1000}\rangle$ --- an MPS obtained by compressing the final state $|\psi_{3000}\rangle$ to $D^* = 1000$. We find that the compression does not affect the results as long as the overlap between the two states is maximized ($ \mathcal{F} = |\langle\phi_{1000}|\psi_{3000}\rangle| \sim 98\%$). At the same time, it substantially reduces the numerical cost of computing the correlation functions, particularly the DSF. The even-sector results correspond to $|\psi_{512}\rangle$. 

As a concluding remark, we note that for highly frustrated systems, such as the one studied in this work, the details of a DMRG simulation can have a strong impact on the resulting variational state. This highlights the importance of carefully assessing convergence and the robustness of the obtained ground states before evaluating physical observables.

 \begin{figure}[t]
\includegraphics[width=\linewidth]{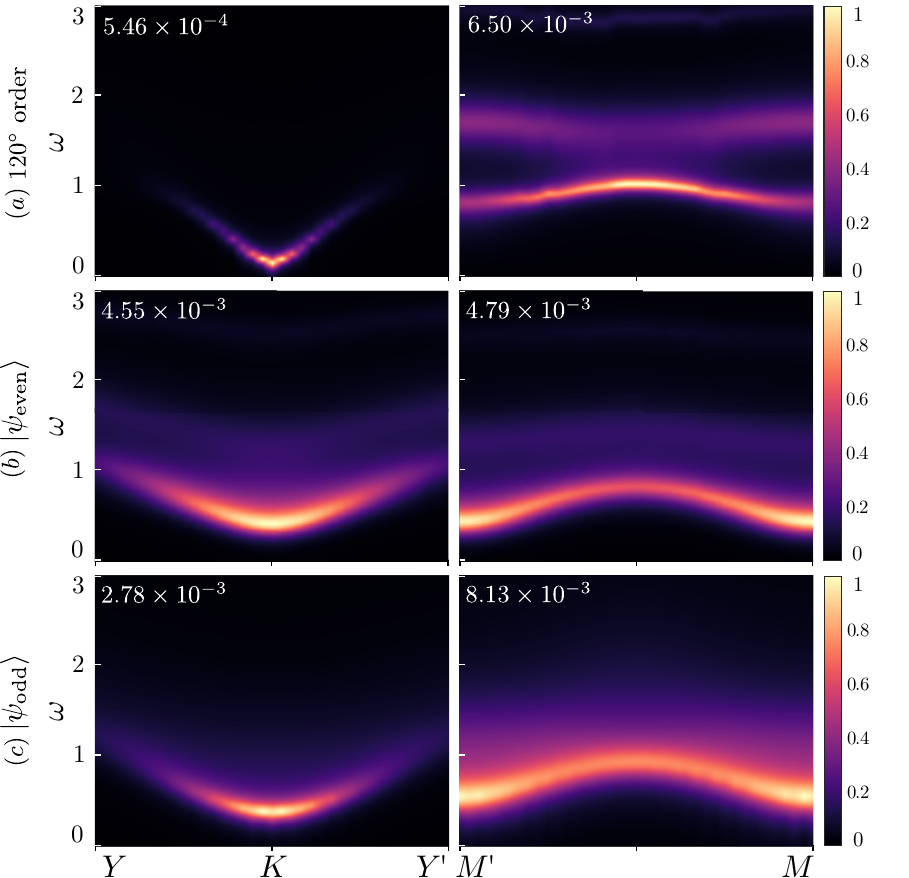} \vspace{-7mm}
\caption{The dynamical structure factor computed for (a) the $120\degree$ ordered phase at $J_2 = 0$  and (b,c) the putative QSL phase at $J_2/J_1 = 0.125$ along the momentum cuts indicated in Fig.~\ref{fig:QSL_static}(e). Each panel is normalized individually to highlight the corresponding spectral profiles. }
\label{fig:DSF_old} 
\vspace{-5mm}
 \end{figure}

\section{Data Processing}
\label{sec:SM_Extra}

The real discrete frequencies and spectral weights $\{\omega_\alpha,S_{\alpha}\}$ obtained with \task \ in our dynamical computations are broadened using continued fraction expansion~(CFE) scheme (see Sec.~S-2 B of the Supplemental Material of Ref.~\onlinecite{Kovalska2025}). For the plots presented in the main text, we set the broadening parameters to $\sigma = 0.4$ and $\mathcal{N}_{\mathrm{CFE}} = 2$. For the cases studied in this work, we find that increasing the number of iterations $\Nkr$ beyond 50 does not lead to quantitative changes in the spectra within the resolution set by our choice of broadening.

 \begin{figure}[t]
 \includegraphics[width=0.8\linewidth]{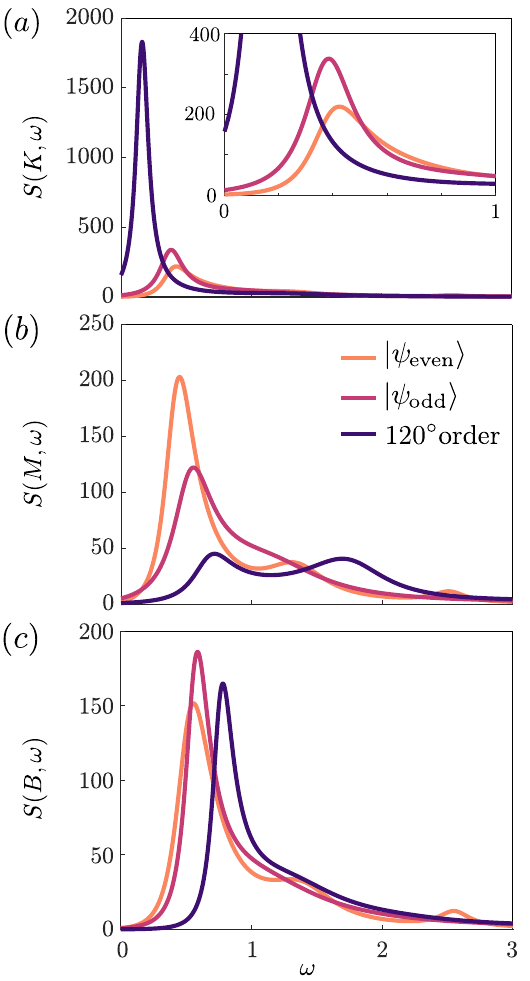}
\caption{The dynamical structure factor computed for the $120\degree$ ordered phase at $J_2 = 0$  and the putative QSL phase at the (a) $K$, (b) $M$ and (c) $B$ points.} 
\label{fig:w_vs_A} 
 \end{figure} 

The broadened data in Fig.~\ref{fig:QSL_DSF} of the main text is normalized to the maximum value $S_{\mathrm{max}}(\vec{\mathrm{k}},\omega)$ among all considered momentum points within a given dataset. This choice allows for an unbiased representation of the spectral weight distribution across the Brillouin zone. However, some of the features in the spectral profile, particularly in the $120\degree$ ordered phase, become difficult to resolve under this global normalization. Figure~\ref{fig:DSF_old} presents the DSF for the $120\degree$ ordered phase [Fig.~\ref{fig:DSF_old}(a)] and QSL candidate phase [Fig.~\ref{fig:DSF_old}(b,c)] with each panel normalized individually to highlight the lineshapes of the emergent spectral features.

In addition, we provide line cuts at $K$ [Fig.~\ref{fig:w_vs_A}(a)], $M$ [Fig.~\ref{fig:w_vs_A}(b)] and $B$ [Fig.~\ref{fig:w_vs_A}(c)] points to allow for comparison of relative peak intensities for the $120\degree$ ordered and putative QSL phases. One notable observation is that, although the intensity of the spectral peak at the $K$ point is larger for \Psiodd \ than for \Psieven, the difference between the two is negligible in comparison to the sharp long-lived quasiparticle found for the ordered phase.

\section{Roton-like Minimum at $Y_1$}
\label{sec:SM_ABcut}

 \begin{figure}[b]
 \vspace{-5mm}
 \includegraphics[width=\linewidth]{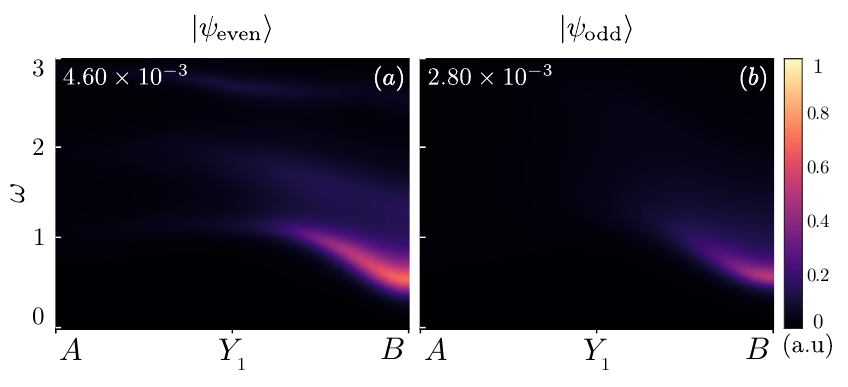} \vspace{-7mm}
\caption{The dynamical structure factor computed along the $A-B$ path for the the putative QSL phase in the even (a) and odd (b) sectors.}
\label{fig:DSF_AB} 
\vspace{-5mm}
 \end{figure} 

\begin{figure*}[t]
\includegraphics[width=\textwidth]{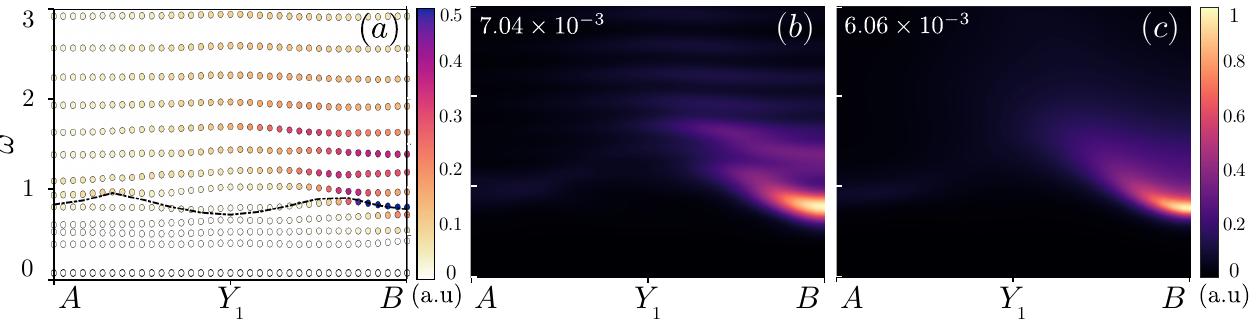}
\caption{The dynamical structure factor computed along the $A-B$ path for the $120^\circ$ ordered phase at $J_2 = 0$: (a) discrete data, (b) Gaussian broadening with $\sigma = 0.1$, and (c) CFE with $N_{\mathrm{CFE}} = 2$ and $\sigma = 0.4$. The dashed line in panel (a) corresponds to the quasiparticle ansatz dispersion line from Ref.~\cite{Drescher2025}.}
\label{fig:AB_NEEL_individual} 
\vspace{-5mm}
 \end{figure*}

Figure~\ref{fig:AB_NEEL_individual} shows the dynamical structure factor for the $120^\circ$ ordered phase, computed along the $A$–$B$ path in the Brillouin zone [see Fig.~\ref{fig:QSL_static}(e)]. Previous works~\cite{Ferrari2019,Drescher2023,Drescher2025} reported the emergence of a roton-like minimum at the $Y_1$ point along this path, similar to those observed at the $M$ and $M'$ points. The appearance of such a feature—referred to as avoided quasiparticle decay—was first proposed in Ref.~\onlinecite{Verresen2019}, where strong renormalizations of the lowest magnon mode were attributed to interactions with the two-magnon continuum.

In Fig.~\ref{fig:AB_NEEL_individual}(a), we compare the discrete \task\ data 
(solid points) with the quasiparticle ansatz dispersion (dashed line) obtained in Ref.~\onlinecite{Drescher2025}. While we find clear agreement near the $B$ point, where most of the spectral weight for this cut is concentrated [see also Fig.~\ref{fig:AB_NEEL_individual}(b,c)], there is no one-to-one correspondence between the tangent-space excitations and the quasiparticle ansatz dispersion. Nevertheless, it is plausible that a physical excitation of the microscopic Hamiltonian~\eqref{eq:HeisenbergHamiltonian} could be represented as a superposition of tangent-space eigenstates.

We also find that the spectral weight near the conjectured minimum is significantly weaker than at the dominant features. Upon broadening the discrete data [Fig.~\ref{fig:AB_NEEL_individual}(a)] using (i) Gaussian broadening with $\sigma = 0.1$ and (ii) CFE broadening with $\sigma = 0.4$ and $\mathcal{N}_{\mathrm{CFE}} = 2$, no resolvable mode is observed along the proposed dispersion line. Instead, the low-energy excitation at the $B$ point appears to merge with the two-particle continuum, reemerging only as a faint line at the $A$ point. Overall, within the resolution of our data, we find no strong indication of a roton-like minimum at the $Y_1$ point for the TLHAF at $J_2=0$. This observation does not rule out its existence, but indicates that any such feature is either very weak or difficult to resolve in our calculations.

For completeness, we include the $A$–$B$ path DSF for \Psieven \ [Fig.~\ref{fig:DSF_AB}(a)] and \Psiodd \ [Fig.~\ref{fig:DSF_AB}(b)], with a global normalization along all considered momentum points [see Fig.~\ref{fig:QSL_static}(e)].

\raggedbottom 

\end{document}